\documentclass [a4paper,fleqn, 12pt]{article}
\usepackage{graphicx}
\usepackage {subfigure,epsfig}

\usepackage {amsmath} \usepackage{amssymb} \usepackage{cite} \usepackage{amsthm}
\numberwithin{equation}{section} \numberwithin{table}{section} \mathindent=0pt
\theoremstyle{plain} \newtheorem{theorem}{Theorem}

\numberwithin{theorem}{section}

\begin{document}

\title{\textbf{Polygons for finding exact solutions of nonlinear differential equations}}
\author{Nikolai A. Kudryashov and Maria V. Demina}
\date{Department of Applied Mathematics\\
Moscow Engineering and Physics Institute\\ (State University)\\
31 Kashirskoe Shosse, 115409, Moscow, \\ Russian Federation}
\maketitle

\begin{abstract}
New method for finding exact solutions of nonlinear differential
equations is presented. It is based on constructing the polygon
corresponding to the equation studied. The algorithms of power
geometry are used. The method is applied for finding one --
parameter exact solutions of the generalized Korteveg -- de Vries --
Burgers equation, the generalized Kuramoto - Sivashinsky equation,
and the fifth -- order nonlinear evolution equation. All these
nonlinear equations contain the term $u^mu_x$. New exact solitary
waves are found.
\end{abstract}



\section{Introduction}

One of the most important problems of nonlinear models analysis is
the construction of their partial solutions. Nowadays this problem
is widely discussed. We know that the inverse scattering transform
\cite{Gardner01, Ablowitz01, Ablowitz03} and the Hirota method
\cite{Hirota01, Ablowitz03, Kudryashov01} are very useful in looking
for the solutions of exactly solvable nonlinear equations, while
most of the nonlinear differential equations describing various
processes in physics, biology, economics and other fields of science
do not belong to the class of exactly solvable equations.

Certain substitutions containing special functions are usually used
for determination of partial solutions of not integrable equations.
The most famous algorithms are the following: the singular manifold
method \cite{Weiss01, Conte01, Choudhary01, Kudryashov02,
Kudryashov03, Kudryashov04, Kudryashov05}, the Weierstrass function
method \cite{Kudryashov05, Kudryashov06, Yan01}, the tanh--function
method \cite{Lou01, Parkes01, Elwakil01, Fan01, Fan02,
Kudryashov07}, the Jacobian elliptic function method \cite{Liu01,
Fu01, Fu02}, the trigonometric function method \cite{Fu03, Yan02}.

Lately it was made an attempt to generalize most of these methods
and as a result the simplest equation method appeared
\cite{Kudryashov08, Kudryashov09}. Two ideas lay in the basis of
this method. The first one was to use an equation of lesser order
with known general solution for finding exact solutions. The second
one was to take into account possible movable singularities of the
original equation. Virtually both ideas existed though not evidently
in some methods suggested earlier.

However the method introduced in \cite{Kudryashov08, Kudryashov09}
has one essential disadvantage concerning with an indeterminacy of
the simplest equation choice. This disadvantage considerably
decreases the effectiveness of the method. In this paper we present
a new method for finding exact solutions of nonlinear differential
equations, which greatly expands the method \cite{Kudryashov08,
Kudryashov09} and which is free from disadvantage mentioned above.
When working out our method we used the ideas of power geometry
recently developed in \cite{Bruno01, Bruno02}. With a help of the
power geometry we show that the search of the simplest equation
becomes illustrative and effective. Also it is important to mention
that the results obtained are sufficiently general and can be
applied not only to finding exact solutions but also to constructing
transformations for nonlinear differential equations.

The paper outline is as follows. New method for finding exact
solution of nonlinear differential equation is introduced in section
2. The application of our approach to the generalized Korteveg -- de
Vries -- Burgers equation is presented in sections 3. Solitary waves
of the generalized Kuramoto -- Sivashinsky equation and of the
nonlinear fifth--order evolution equation are found in sections 4
and 5, accordingly.

\section{Method applied}

Let us assume that we look for exact solutions of the following
nonlinear n-order ODE
\begin{equation}
\label{e:2.1}M_n\left(y(z),y_z(z),y_{zz}(z),\ldots ,z\right)=0.
\end{equation}
In power geometry any differential equation is regarded as a sum of
ordinary and differential monomials. Every monomial can be
associated with a point on the plane according to the following
rules
\begin{equation}\label{e:2.2}
C_2z^{q_1}y^{q_2}$ $\longrightarrow (q_1,q_2), \,\,\,C_2d^ky/dz^k
\longrightarrow (-k,1).
\end{equation}
Here $C_1,C_2$ are arbitrary constants. When monomials are
multiplied their coordinates are added. The set of points
corresponding to all monomials of a differential equation forms its
carrier. Having connected the points of the carrier into the convex
figure we obtain a convex polygon called a polygon of a differential
equation. Thus the nonlinear ODE \eqref{e:2.1} is associated with a
polygon $L_1$ on the plane.

Now let us assume that a solution $y(z)$ of the basic equation can
be expressed through solutions $Y(z)$ of another equation. The
latter equation is called the simplest equation. Equations having
known general solutions or solutions without movable critical points
are usually taken as the simplest equations. Consequently we have a
relation between $y(z)$ and $Y(z)$
\begin{equation}\label{e:2.3}
y(z)=F(Y(z),Y_z(z),\ldots ,z).
\end{equation}
The main problem is to find the simplest equation. Substitution
\eqref{e:2.3} into the basic equation \eqref{e:2.1} yields
transformed differential equation which is characterized by new
polygon $L_2$. The suitable simplest equations should be looked for
among the equations whose polygons, first, are of lesser or equal
area than $L_2$ and, second, have all or certain part of edges
parallel to those of $L_2$. Let us suppose that we have found such
polygon $L_3$. Then we can write out the simplest equation
\begin{equation}\label{e:2.4}
E_m(Y(z),Y_z(z),\ldots ,z)=0.
\end{equation}
It is important to mention that the choice of the simplest equation
is not unique. If the following relation
\begin{equation}\label{e:2.5}
M_n(F(Y,Y_z,\ldots ,z))=\hat{R}\,E_m(Y,Y_z,\ldots ,z)
\end{equation}
(where $\hat{R}$ is a differential operator) is true then it means
that for any solution $Y(z)$ of the simplest equation \eqref{e:2.4}
there exists a solution of \eqref{e:2.1}.

In this paper we will consider a case of identity substitution, i.e.
\begin{equation}
\label{e:2.6}
y(z) \equiv Y(z).
\end{equation}

Then we should study the polygon $L_1$ of the equation \eqref{e:2.1}
and try to find the polygon $L_3$ generated by the simplest equation
\eqref{e:2.4}. In this case, our method can be subdivided into four
steps.

\textit{The first step.} Construction of the polygon $L_1$, which
corresponds to the equation studied.

\textit{The second step.} Construction of the polygon $L_3$ which
characterizes the simplest equation. This polygon should posses
qualities discussed above.

\textit{The third step.} Selection of the simplest equation with
unknown parameters in such a way that it generates the polygon
$L_3$.

\textit{The fourth step.} Determination of the unknown parameters in
the simplest equation.

To demonstrate our method application let us find exact solutions of
nonlinear differential equations belonging to the following class
\begin{equation}
\label{e:2.7}\sum_{k=1}^{n}\,a_{k}\,\frac{d^{k}y(z)}{dz^k}-C_0\,y(z)+\frac{\alpha}{m+1}\,y^{m+1}(z)=0.
\end{equation}

This class of nonlinear ordinary differential equations containes a
number of differential equations that corresponds to the famous
nonlinear evolution equations. They are the Burgers -- Huxsley
equation, the Korteveg -- de Vries -- Burgers equation, the Kuramoto
-- Sivashinsky equation and so on. The following points
$M_1=(-1,1)$, $M_2=(-2,1)$, $\ldots$, $M_n=(-n,1)$,
$M_{n+1}={(0,1)}$, and $M_{n+2}=(0,m+1)$ are assigned to the
monomials of the equations \eqref{e:2.7}. The polygon $L_1$
corresponding to the equations \eqref{e:2.7} is the triangle
presented at figure 1 ((a): $m>0$ and (b): $m<0$).

\begin{figure}[h]
 \centerline{
 \subfigure[]{\epsfig{file=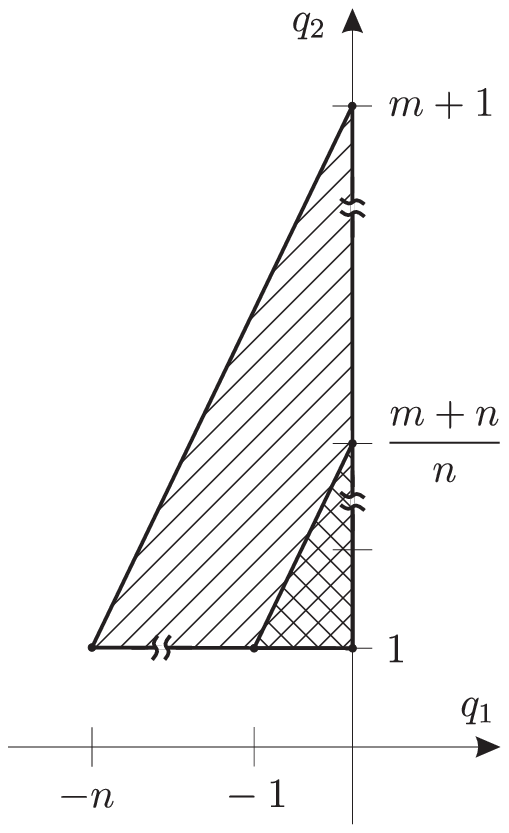,width=60mm}\label{fig1:z_post_a}}
 \subfigure[]{\epsfig{file=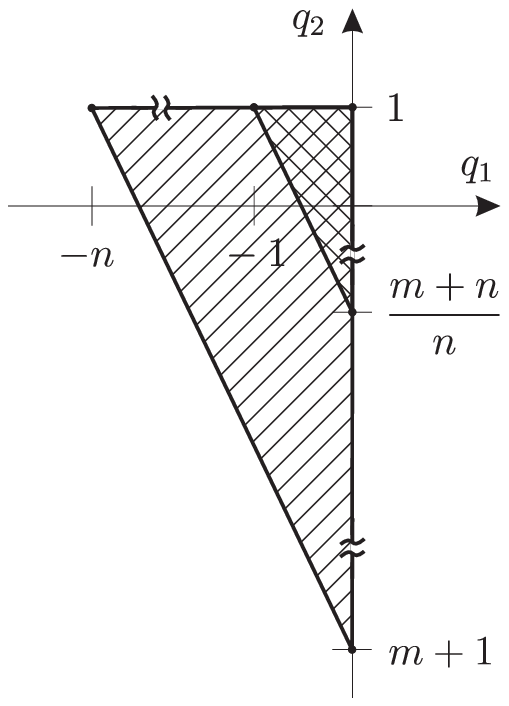,width=60mm}\label{fig1:z_post_b}}
 }
 \caption{Polygons corresponding to the equations studied and to the simplest equations.}\label{fig1:z_post}
\end{figure}

Examining the triangle $L_1$ we can easily find the polygon $L_3$
assigned to the simplest equation of the first order (see smaller
triangles at figure 1). Hence the simplest equation can be written
as
\begin{equation}
\label{e:2.8}y_z(z)=A\,y^{\frac{n+m}{n}}(z)\,+\,B\,y(z).
\end{equation}
This equation can be transformed to a linear equation. Its general
solution takes the form
\begin{equation}
\label{e:2.9}y(z)\,=\,\left(C_1\,\exp{\left\{-\frac{B\,m\,z}
{n}\right\}-\,\frac{A}{B}}\right)^{-\frac{n}{m}}.
\end{equation}

Later, having found the simplest equation, we should determine the
coefficients $A$ and $B$ for members of the class \eqref{e:2.7}.
Then exact solutions of these equations will be expressed through
the general solution of the simplest equation.

\section{Exact solutions of the generalized Korteveg -- de Vries -- Burgers equation}

Let us find exact solutions of the generalized Korteveg -- de Vries
-- Burgers equation with a help of our method. This equation can be
written as

\begin{equation}
\label{e:3.1}u_t\,-\alpha
\,u^{m}\,u_x\,+\,\beta\,u_{xxx}=\nu\,u_{xx}.
\end{equation}

At $m=1$ and $\nu=0$ \eqref{e:3.1} is the famous Korteveg -- de
Vries equation. Cauchy problem for this equation is solved by the
inverse scattering transform \cite{Gardner01}. Soliton solutions of
this equation were found using the Hirota method \cite{Hirota01}. At
$m=1$ and $\nu\neq0$ a dissipation is taken into account. In this
case, the equation \eqref{e:3.1} is not integrable one. Its special
solutions were obtained in the work \cite{Kudryashov02}. Later these
solutions were rediscovered many times.

Using the travelling wave reduction
\begin{equation}\label{e:3.2}
u(x,t)=y(z),\,\,\,\,\,\,z=x-C_0\,t
\end{equation}
and integrating with respect to $z$, we get
\begin{equation}\label{e:3.3}
\beta\,y_{zz}-\nu\,y_z\,-\,C_0\,y\,-\,\frac{\alpha}{m+1}\,\,y^{m+1}\,=0,
\end{equation}
where the constant of integration is equated to zero.

Taking into account the results obtained in section 2, we can write
the simplest equation in the form
\begin{equation}
\label{e:3.4}y_{z}(z) =A\,y^{\frac{m+2}{2}}(z)+B\,y(z).
\end{equation}

 The general solution of the equation \eqref{e:3.4} is
\begin{equation}
\label{e:3.4a}y{(z)}\,=\,\left(C_1\,\exp{\left\{-\frac{B\,m\,z}
{2}\right\}-\,\frac{A}{B}}\right)^{-\frac{2}{m}}.
\end{equation}

Let us formulate our results in the form of the following theorem.

\begin{theorem}
\label{T:3.1.} Let $y(z)$ be a solution of the equation

\begin{equation}
\label{e:3.8}y_z(z)\,=\,\pm\,{\frac {\sqrt{2\,\alpha}}{\sqrt
{\beta\,({m}+2)(m+1)}}}\,y^{\frac{m+2}{2}}(z)+{\frac {2\,\nu}{
\beta\, \left( m+4 \right)}}\,y(z)
\end{equation}
Then there exists a solution of the equation \eqref{e:3.3} that
coincides with $y(z)$ provided that $m\neq0$, $m\neq-1$, $m\neq-2$,
$m\neq-4$, and
\begin{equation}
\label{e:3.9}C_0\,=-\,\frac{2\,\nu^2(m+2)}{\beta\,(m+4)^2}.
\end{equation}

\end{theorem}

\begin{proof} Let $M(y,y_z,y_{zz})=0$ be the generalized Korteveg -- de Vries --
Burgers equation \eqref{e:3.3}. By $E(y,y_z)=0$ denote the equation
\eqref{e:3.8}. Substituting
\begin{equation}
\label{e:3.10}
y_z(z)\,=\,E(y,y_z)\,+\,A\,y^{\frac{m+2}{2}}(z)\,+B\,y(z)
\end{equation}
into the equation \eqref{e:3.3} and equating expressions at
different powers of  $y(z)$ to zero we get algebraic equations for
parameters $A$, $B$, and $C_0$ in the form

\begin{equation}
\label{e:3.11}\,\beta\,{A}^{2} \left( m+2 \right) \left( m+1 \right)
-2\,\alpha =0,
\end{equation}

\begin{equation}
\label{e:3.12} \left( m+1 \right)  \left( \beta\,{B}^{2}-\nu\,B-{\it
C_0}
 \right)=0,
\end{equation}

\begin{equation}
\label{e:3.13}A \left( m+1 \right)  \left(
\beta\,B\,m+4\,\beta\,B-2\,\nu
 \right)=0.
\end{equation}
Equations \eqref{e:3.11}, \eqref{e:3.12} and \eqref{e:3.13} can be
solved and we obtain

\begin{equation}
\label{e:3.14}A_{1,2}=\pm\,{\frac {\sqrt{2\,\alpha}}{\sqrt
{\beta\,({m}+2)(m+1)}}},
\end{equation}

\begin{equation}
\label{e:3.15}C_{{0}}=- B\, \left( \nu-\beta\,B \right),
\end{equation}

\begin{equation}
\label{e:3.16}B={\frac {2\,\nu}{ \beta\, \left( m+4 \right)}}.
\end{equation}

Consequently we have found the parameters $A$,  $B$, $C_0$ of the
theorem and the relation
\begin{equation}
\label{e:3.17}M(y,y_z,y_{zz})\,=\hat{R}\,E(y,y_z),
\end{equation}
where $\hat{R}$ is a differential operator.
\end{proof}

Thus the solution of the equation \eqref{e:3.3} is found. It takes
the form

\begin{equation}
\label{e:3.18}y(z)\,=\,\left(\frac {\sqrt{\alpha
\,\beta}\,(m+4)}{\sqrt{2\,\nu^2\,(m+2)(m+1)}}+C_1\,\exp{\left\{-\frac{\nu\,m\,z}
{\beta\,(m+4)}\right\}}\right)^{-\frac{2}{m}},
\end{equation}
where $C_1$ is an arbitrary constant.

Note that in the case, $m\rightarrow\,0$ the value of solitary wave
velocity \eqref{e:3.9} tends to
$C_0\,=\,-{(4\,\nu^2)}/{(16\,\beta)}$, but in the case,
$m\rightarrow\infty$ we get $C_0\rightarrow\,0$.

It can be shown the correctness of the following formula

\begin{equation}
\label{e:3.19}\left(S_0+C_1\,\exp{\left \{-\,k\,z\,\right
\}}\right)^{-p}\,=\,(2\,S_0)^{-p}\left(1+\tanh{\left\{\frac{k\,(z+\varphi_0)}{2}\right\}}\right)^{p},
\end{equation}
where $S_0$ is an arbitrary constant and $\varphi_0$ is a constant
connected with the coefficient $C_1$ by the relation

\begin{equation}
\label{e:3.20}\varphi_0=\frac 1k\,\ln{\frac{C_1}{S_0}}.
\end{equation}

Using \eqref{e:3.19} the solution \eqref{e:3.18} can be rewritten as

\begin{equation}
\label{e:3.21}y(z)\,=\,\left(\frac{\nu^2\,(m+1)(m+2)}{2\,\alpha\,\beta\,(m+4)^2}\right)^{\frac
1m}\,\left(1+\tanh{\left\{\frac{\nu\,m\,(z+\varphi_0)}{2\,\beta\,(m+4)}\right\}}\right)^{\frac2m}.
\end{equation}

Assuming $m=1$ in \eqref{e:3.21}, we get a solution of the Korteveg
-- de Vries -- Burgers equation in the form

\begin{equation}
\label{e:3.22}y(z)\,=\frac{3\,\nu^2}{25\,\alpha\,\beta}\left(1+\tanh{\left\{\frac{\nu\,(z+\varphi_0)}{10\,\beta\,}\right\}}\right)^{2}.
\end{equation}
This solution was found in \cite{Kudryashov02}.

\section{Exact solutions of the generalized Kuramoto -- Sivashinsky equation}

Let us look for exact solutions of the generalized Kuramoto --
Sivashinsky equation, which can be written as

\begin{equation}\begin{gathered}
\label{e:5.0}u_t\,+\,\alpha\,u^{m}\,u_x\,+\,\delta\,u_{xx}+\,\beta\,u_{xxx}+\,\gamma\,u_{xxxx}\,=\,0.
\end{gathered}\end{equation}

Equation \eqref{e:5.0} at $m=1$ is the famous Kuramoto --
Sivashinsky equation \cite{Kuramoto01, Sivashinsky01}, which
describes turbulent processes. Exact solutions of the equation
\eqref{e:5.0} at $\beta=0$ were first found  in \cite{Kuramoto01}.
Solitary waves of  \eqref{e:5.0} at $\beta\neq0$ were obtained in
\cite{Kudryashov02} and periodical solutions of this equation at
$\beta\neq0$ were first presented in \cite{Kudryashov05}.

Using the variables

\begin{equation}\begin{gathered}
\label{e:5.0a}
x'\,=\,x\,\sqrt{\frac{\delta}{\gamma}},\,\,\,\,\,\,t'=t\,\frac{\delta^2}{\gamma},\,\,\,\,\,\,u'=u,\,\,\,\,\,\,
\sigma\,=\,\frac{\beta}{\sqrt{\gamma\,\delta}}
,\,\,\,\,\,\,\alpha'\,=\,\frac{\alpha\,\sqrt{\delta\,\gamma}}{\delta^2},
\end{gathered}\end{equation}

we get an equation in the form (the primes are omitted)

\begin{equation}\begin{gathered}
\label{e:5.1}u_t\,+\,\alpha\,u^{m}\,u_x\,+\,u_{xx}+\,\sigma\,u_{xxx}+\,u_{xxxx}\,=\,0.
\end{gathered}\end{equation}

Taking into account the travelling wave reduction
\begin{equation}\label{e:5.2}
u(x,t)=y(z),\,\,\,\,\,\,z=x-C_0\,t
\end{equation}
and integrating with respect to $z$, we get the equation
\begin{equation}\label{e:5.5}
y_{zzz}\,+\sigma\,y_{zz}\,+\,y_{z}\,-\,C_0\,y\,+\,\frac{\alpha}{m+1}\,\,y^{m+1}\,=0.
\end{equation}
Here a constant of integration is equated to zero.

In this case, the simplest equation is the following

\begin{equation}
\label{e:5.6}y_{z} =A\,y^{\frac{m+3}{3}}+B\,y,
\end{equation}
where $A$ and $B$ are parameters to be found.

The general solution of \eqref{e:5.6} takes the form
\begin{equation}
\label{e:5.6a}y(z) =\,\left(C_1\,\exp{\left\{-\frac{B\,m\,z}
{3}\right\}-\,\frac{A}{B}}\right)^{-\frac{3}{m}}.
\end{equation}

Let us present our result in the following theorem.

\begin{theorem}
\label{T:5.1.} Let $y(z)$ be a solution of the equation

\begin{equation}
\label{e:5.7}y_{z}(z)\,=\left(-\frac{9\,\alpha}{2\,{m}^{3}+11\,{m}^{2}+18\,m+9}\right)^{\frac
13}\,y^{\frac{m+3}{3}}(z)\,\mp\,{\frac {3}{\sqrt
{2\,{m}^{2}+18\,m+27}}}\,\,y(z).
\end{equation}
Then $y(z)$ is a solution of the equation \eqref{e:5.5} in the case,
$m\,\neq\,0$, $m\,\neq\,-1$, $m\,\neq\,-\frac {3}{2}$,
$m\,\neq\,-3$, $m\,\neq\,\frac{-9\,\pm\,3\,\sqrt{3}}{2}$, and
\begin{equation}
\label{e:5.8}C_{{0}}=\mp\,{\frac {3\,(2\,{m}^{2}+9\,m+9)}{ \left(
2\,{m}^{2}+18\,m+27 \right) ^{3/2}}},\,\,\,\,\,\sigma=\pm\,{\frac
{3\,(3+m)}{\sqrt {2\,{m}^{2}+18\,m+27}}}.
\end{equation}

\end{theorem}

\begin{proof} Let $M(y,y_z,y_{zz})=0$ be the generalized Kuramoto -- Sivashinsky equation \eqref{e:5.5}.
By $E(y,y_z)=0$ denote the the equation \eqref{e:5.6}. Substituting
\begin{equation}
\label{e:5.9}
y_z(z)\,=\,E(y,y_z)\,+\,A\,y^{\frac{m+3}{3}}(z)\,+\,B\,y(z)
\end{equation}
into equation \eqref{e:5.5} and equating coefficients at powers of
$y(z)$ to zero, yields algebraic equations for parameters $A$, $B$,
$C_0$ and $\sigma$ in the form

\begin{equation}
\label{e:5.10} {A}^{3} \left( 2\,m+3 \right)  \left(m+ 3 \right)
\left( m+1 \right) +9\,\alpha=0,
\end{equation}

\begin{equation}
\label{e:5.11} \left( m+1 \right)  \left(
{B}^{3}+\sigma\,{B}^{2}+B-{\it C_0}
 \right)\,=\,0,
\end{equation}

\begin{equation}
\label{e:5.12}{A}^{2} \left( 3+m \right)  \left( m+1 \right) \left(
B\,m+3 \,B+\sigma \right) =0,
\end{equation}

\begin{equation}
\label{e:5.13}\left( m+1 \right)  \left(
2\,{\sigma}^{2}{m}^{2}+18\,{\sigma}^{2}\,m\,+\,27\,{\sigma}^{2}\,-\,9\,{m}^{2}\,-\,54\,m\,-\,81
\right)=0.
\end{equation}

Solutions of the equations \eqref{e:5.10}, \eqref{e:5.11},
\eqref{e:5.12} and \eqref{e:5.13} can be written as

\begin{equation}\begin{gathered}
\label{e:5.14}A_1\,=(-9\,\alpha)^{1/3}\,
(\,{m}^{3}+11\,{m}^{2}+18\,m+9)^{-1/3},\\
\\
A_{2,3}\,=(-9\,\alpha)^{1/3}\,\left(-\,\frac12\,\mp\,i\,\frac{\sqrt
{ 3}}{2}\right)\, \left({m}^{3}+11\,{m}^{2}+18\,m+9\right)^{-1/3},
\end{gathered}\end{equation}

\begin{equation}
\label{e:5.16}C_0\,=\,B\, \left( {B}^{2}+\sigma\,B+1 \right)
,\,\,\,\,\,\,\,B\,=-{\frac {\sigma}{3+m}},
\end{equation}

\begin{equation}
\label{e:5.17}\sigma=\pm\,{\frac {3\,(3+m)}{\sqrt
{2\,{m}^{2}+18\,m+27}}}.
\end{equation}

In final expressions we will take into consideration $A_1$ only.

So we have found parameters of the equation \eqref{e:5.6a},
conditions \eqref{e:5.8}, and the relation
\begin{equation}
\label{e:5.18}M(y,y_z,y_{zz})\,=\hat{R}\,E(y,y_z),
\end{equation}
where $\hat{R}$ is a differential operator. This completes the
proof.
\end{proof}

The general solution of the equation \eqref{e:5.8} can be presented
in the form

\begin{equation}
\label{e:5.19}y(z)\,=\,\left(C+C_1\,\exp{\left\{\pm{\frac{\,m\,z}{\sqrt
{2\,{m}^{2}+18\,m+27}}}\right\}}\right)^{-\frac{3}{m}},
\end{equation}
where $C_1$ is an arbitrary constant and $C$ is determined by
expression

\begin{equation}\label{e:5.20}
C=\,\pm\,{\frac {\sqrt [3]{-9\,\alpha}\,\sqrt
{2\,{m}^{2}+18\,m+27}}{3\sqrt [3]{ \left( m+3 \right)
 \left( 2\,{m}^{2}+5\,m+3 \right)} }}.
\end{equation}

Again we note that the value of solitary wave velocity
\eqref{e:5.19} tends to $C_0\,=\,\mp\,1/\sqrt{27}$ as
$m\rightarrow\,0$, but at the same time $C_0\rightarrow\,0$ as
$m\rightarrow\infty$.

Assuming $m=1$ in \eqref{e:5.19}, we obtain the known solitary wave

\begin{equation}\label{e:5.21}y(z)=\left(\frac{\sqrt
{47}\,\sqrt
[3]{-225\,\alpha}}{30}+C_1\exp{\left\{\pm{\frac{\,x\,-\,C_0\,t}{\sqrt
{47}}}\right\}}\right)^{-{3}},\,\,\,C_0=\mp\frac{60}{47\sqrt{47}}.
\end{equation}
and the value of the parameter $\sigma$: $\sigma\,=\,\pm
12/\sqrt{47}$. This solution of the Kuramoto -- Sivashinsky equation
is the kink \cite{Kudryashov02, Kudryashov03, Kudryashov04,
Kudryashov05, Kudryashov06}.

\section{Exact solutions of the fifth -- order nonlinear evolution equation}

Consider the fifth--order nonlinear evolution equation of the form

\begin{equation}\begin{gathered}
\label{e:6.1}u_t\,-\alpha\,u^{m}\,u_x\,+\,\sigma\,u_{xx}+\,\delta\,u_{xxx}+\,\beta\,u_{xxxx}\,+\,u_{xxxxx}=\,0.
\end{gathered}\end{equation}

It has the travelling wave solution
\begin{equation}\label{e:6.2}
u(x,t)=y(z),\,\,\,\,\,\,z=x-C_0\,t,
\end{equation}
where $y(z)$ satisfies
\begin{equation}\label{e:6.5}
y_{zzzz}+\beta\,y_{zzz}\,+\delta\,y_{zz}\,+\sigma\,y_{z}\,-\,C_0\,y\,-\,\frac{\alpha}{m+1}\,\,y^{m+1}\,=0.
\end{equation}

Without loss of generality it can be set $\alpha=m+1$. The simplest
equation for \eqref{e:6.5} is the following
\begin{equation}
\label{e:6.6}y_{z}(z) =A\,y^{\frac{m+4}{4}}(z)+B\,y(z),
\end{equation}
where $A$ and $B$ are parameters that should be found.

The general solution of \eqref{e:6.6} is
\begin{equation}
\label{e:6.6a}y_{z} =\,=\,\left(C_1\,\exp{\left\{-\frac{B\,m\,z}
{4}\right\}-\,\frac{A}{B}}\right)^{-\frac{4}{m}}
\end{equation}

Let us summarize our results in the theorem.

\begin{theorem}
\label{T:6.1.} Let $y(z)$ be a solution of the equation

\begin{equation}
\label{e:6.9a}y_{z}(z)\,=\,{\frac { {2}\sqrt[4]{2}}{\sqrt
[4]{(m+4)(3m+4)(m+2)}}}\,\,y^{\frac{m+4}{4}}(z)\,-\,{\frac
{2\,\beta}{3\,m+8}}\,\,y(z).
\end{equation}
Then $y(z)$ is also a solution of the equation \eqref{e:6.5}
provided that $m\,\neq\,0$, $m\,\neq\,-1$, $m\,\neq\,-\frac{4}{3}$,
 $m\,\neq\,-2$, $m\,\neq\,-\frac {8}{3}$, $m\,\neq\,-4$, and

\begin{equation}
\label{e:6.9b}C_{{0}}\,=\,-{\frac {{\beta}^{4} (m+4)(3m+4)(m+2)}{
2\,\left( 3\,m+8 \right) ^{4}}},
\end{equation}

\begin{equation}
\label{e:6.9c}\delta=\,{\frac {{\beta}^{2} \left(
11\,{m}^{2}+72\,m+96 \right) }{4\,
 \left( 3\,m+8 \right) ^{2}}},\,\, \,\,\,\, \sigma\,=\,{\frac {{\beta}^{3} \left( {m}^{2}+12\,m+16
\right) }{4\,
 \left( 3\,m+8 \right) ^{2}}}.
\end{equation}

\end{theorem}

\begin{proof} Let $M(y,y_z,y_{zz})=0$ be the equation \eqref{e:6.5}.
By $E(y,y_z)=0$ denote the equation \eqref{e:6.9a}. Substituting

\begin{equation}
\label{e:6.10}
y_z(z)\,=\,E(y,y_z)\,+\,A\,y^{\frac{m+4}{4}}(z)\,+\,B\,y(z)
\end{equation}
into $M(y,y_z,y_{zz})=0$ and equating expressions at different
powers $y(z)$ to zero, we obtain

\begin{equation}
\label{e:6.11}{}\,{A}^{4} \left( m+4 \right)  \left( 3\,m+4 \right)
 \left( m+2 \right) -32=0,
\end{equation}

\begin{equation}
\label{e:6.12}{C_0}-{B}^{4}-\beta\,{B}^{3}-\delta\,{B}^{2}-\sigma\,B=0,
\end{equation}

\begin{equation}
\label{e:6.13} \left( m+4 \right)  \left( m+2 \right)  \left(
3\,mB+2\,\beta+ 8\,B \right)=0,
\end{equation}

\begin{equation}
\label{e:6.14}{}\,\left( m+4 \right)  \left(
7\,{m}^{2}{B}^{2}+48\,m\,{B}^{2}+96\,{B}^{2}+
12\,\beta\,m\,B+48\,\beta\,B+16\,\delta
 \right) =0,
\end{equation}

\begin{equation}\begin{gathered}
\label{e:6.15}{\frac
{9}{64}}\,A\,{m}^{3}{B}^{3}+\frac94\,A\,{m}^{2}{B}^{3}+ {\frac
{27}{2}}\,A\,m\,{B}^{3}+36\,A\,{B}^{3}+{\frac
{9}{16}}\,\beta\,A\,{m}^{2}\,{B}^{2}+\\
\\
+{\frac {27}{4}}\,\beta\,A\,m\,{B}^ {2}+27\,\beta \,A\,{B}^{2}+\frac
94\,\delta\,A\,m\,B +18\,\delta\,A\,B+9\,\sigma\,A=0.
\end{gathered}\end{equation}

Solutions of the equations \eqref{e:6.11}, \eqref{e:6.12},
\eqref{e:6.13},  \eqref{e:6.14}, and \eqref{e:6.15} can be written
in the form

\begin{equation}
\label{e:6.16}A_{1,2}\,=\,\pm\,{\frac {2\,\sqrt {2}}{\sqrt
[4]{2(m+4)(3m+4)(m+2)}}},
\end{equation}

\begin{equation}\begin{gathered}
\label{e:6.17}A_{3,4}\,=\mp\,{\frac {4}{\sqrt {-2\,\sqrt
{2(m+4)(3m+4)(m+2)}}}},
\end{gathered}\end{equation}

\begin{equation}
\label{e:6.18}C_{{0}}=\beta\,{B}^{3}+\delta\,{B}^{2}+\sigma\,B+{B}^{4},
\end{equation}

\begin{equation}
\label{e:6.19}B=-\,{\frac {2\,\beta}{3\,m+8}},
\end{equation}

\begin{equation}
\label{e:6.20}\delta=-{\frac
{7}{16}}\,{m}^{2}{B}^{2}-3\,m{B}^{2}-6\,
{B}^{2}-\frac34\,\beta\,m\,B-3\,\beta\,B,
\end{equation}

\begin{equation}\begin{gathered}
\label{e:6.21}\sigma=-{\frac {1}{64}}\,{m}^{3
}{B}^{3}-\frac14\,{m}^{2}\,{B}^{3}-\frac32\,m\,{B}^{3}-4\,{B}^{3}-\frac{1}{16}\,\beta\,{m}^{2}\,{B}^{2}-\\
\\
-\frac34\,\beta\,m\,{B}^{2}-3\,\beta\,{B}^{2} -\frac14\,\delta\,
m\,B -2\,\delta\,B.
\end{gathered}\end{equation}

In final expressions $A_1,A_2$ will be used only.

From obtained expressions \eqref{e:6.18}, \eqref{e:6.19},
\eqref{e:6.20}, and \eqref{e:6.21} we get conditions \eqref{e:6.9b},
\eqref{e:6.9c}, and the relation
\begin{equation}
\label{e:6.22}M(y,y_z,y_{zz})\,=\hat{R}\,E(y,y_z),
\end{equation}
where $\hat{R}$ is a differential operator.

\end{proof}

The general solution of \eqref{e:6.9a} can be written as

\begin{equation}
\label{e:6.22}y(z)=\left(\pm{\frac {\sqrt [4]{2} \left( 3\,m+8
\right) }{\sqrt
[4]{\beta\,(m+4)(3m+4)(m+2)}}}+C_1\exp{\left\{{{{\frac
{\beta\,mz}{2(3\,m+8)}}}}\right\}}\right)^{-\frac{4}{m}},
\end{equation}
where $C_1$ is an arbitrary constant.

Let us note that in the case, $m\rightarrow\,0$ the value of
solitary wave velocity \eqref{e:6.22} tends to
$C_0\,=\,-\beta^2/256$, but $C_0\rightarrow\,0$ as
$m\rightarrow\infty$.

\section {Conclusion}
In this paper a new method to look for exact solutions of nonlinear
differential equations is presented. Our goal was to express
solutions of equation studied through general solutions of the
simplest equations. The basic idea of our approach is to use
geometrical representations for nonlinear differential equations.
The introduced method allows one to find the suitable simplest
equations. Consequently our method is more powerful than many other
methods that use an a priori expressions for $y(z)$. In our
terminology these methods take the a priori simplest equation. With
a help of our approach we have found exact solutions of the
following nonlinear differential equations: the generalized Korteveg
-- de Vries -- Burgers equation, the generalized Kuramoto --
Sivashinsky equation and nonlinear fifth -- order differential
equation. All these equations belong to the class \eqref{e:2.7}.
Finally we would like to mention that this class can be also
expanded. Exact solutions of equations belonging to the new class
can be constructed with a help of our method.

\section {Acknowledgments}

This work was supported by the International Science and Technology
Center under Project No. B 1213.


\begin{thebibliography}{99}

\bibitem{Gardner01} \textit{C.S. Gardner, J.M. Greene, M.D. Kruskal,
 R.R. Miura},  Phys. Rev. Lett., \textbf{19} (1967) 1095 - 1097

\bibitem{Ablowitz01} \textit{M.J. Ablowitz, D.J. Kaup, A.C. Newell, H. Segur}, Phys. Rev. Lett., \textbf{31} (1973) 125

\bibitem{Ablowitz03} \textit{M.J. Ablowitz , P.A. Clarcson }, Solitons, Nonlinear Evolution Equations and Inverse Scattering,
Cambridge University Press, (1991), 516 p.

\bibitem{Hirota01} \textit{R. Hirota}, Phys. Rev. Lett., \textbf{27} (1971) 1192 - 1194

\bibitem{Kudryashov01} \textit{N.A. Kudryashov}, Analytical theory of
nonlinear differential equations,  Institute of Computer
Investigations, Moscow-Izhevsk, (2004), 360 p. (in Russian).



\bibitem{Weiss01} \textit{J. Weiss ,  M. Tabor, G. Carnevalle},
J. Math. Phys., \textbf{24}, (1983), 522

\bibitem{Conte01} \textit{R. Conte  and M. Musette}, J. Phys.
A.: Math. Gen. \textbf{22}, (1989),  169-177

\bibitem{Choudhary01} \textit{S.R. Choudhary}, Phys Lett. A., \textbf{159}, (1991), 311-317

\bibitem{Kudryashov02} \textit{N.A. Kudryashov},  Journal of
Applied Mathematics and Mekhanics, \textbf{52}, (1988),361-365

\bibitem{Kudryashov03}\textit{N.A. Kudryashov},  Reports of USSR Academy Siences ,  No.2, \textbf{308}, (1989), pp. 294 --
298 (in Russian)

\bibitem{Kudryashov04} \textit{N.A. Kudryashov}, Phys Lett. A., \textbf{155}, (1991), 269 -- 275

\bibitem{Kudryashov05} \textit{N.A. Kudryashov}, Phys Lett. A., \textbf{147}, (1990), 287 -- 291

\bibitem{Kudryashov06} \textit{N.A. Kudryashov},  Journal of Applied Mathematics and Mekhanics,
\textbf{54}, (1990), 372-376

\bibitem{Yan01} \textit{Z.Y. Yan}, Chaos, Solitons and Fractals \textbf{21}, (2004), 1013

\bibitem{Lou01} \textit{S.Y. Lou, G. Huang, H. Ruan},, J. Phys. A.: Math. Gen., \textbf{24}, (1991),  587-590

\bibitem{Parkes01} \textit{E.J. Parkes,  B.R. Duffy}, Computer Physics Communications, \textbf{98}, (1996),  288-300

\bibitem{Elwakil01} \textit{S.A. Elwakil, S.K. El-labany , M.A. Zahran,  R. Sabry}, Phys. Lett. A., \textbf{299}, (2002), 179 -188

\bibitem{Fan01} \textit{E.G. Fan}, Phys. Lett. A., \textbf{277}, 4-5, (2000), 212 -- 218

\bibitem{Fan02} \textit{E.G. Fan}, Phys. Lett. A., \textbf{282}, 1 -- 2, (2002), 18 - 22

\bibitem{Kudryashov07} \textit{N.A. Kudryashov,  E.D. Zargaryan},,  J. Phys. A. Math. and
Gen. \textbf{29}, (1996), 8067-8077

\bibitem{Liu01} \textit{G.T. Liu, T.Y. Fan}, Phys. Lett. A., \textbf{345}, (2005), 161-166

\bibitem{Fu01} \textit{S.K. Liu, Z.T. Fu, S.D. Liu, Q. Zhao}, Phys. Lett. A., \textbf{289}, (2001), 69-74

\bibitem{Fu02} \textit{Z. Fu, L. Zhang, S. Liu, S. Liu}, Phys. Lett. A., \textbf{325}, (2004), 363 -- 369

\bibitem{Fu03} \textit{Z. Fu, S. Liu, S. Liu}, Phys. Lett. A., \textbf{326}, (2004), 364 -- 374

\bibitem{Yan02} \textit{C. T. Yan}, Phys. Lett. A., \textbf{224}, (1996),  77

\bibitem{Kudryashov08} \textit{N.A. Kudryashov}, Phys Lett. A., \textbf{342}, (2005),
99 -- 106

\bibitem{Kudryashov09}\textit{N.A. Kudryashov}, Chaos, Solitons and Fractals, \textbf{24}, (2005),
1217 -- 1231

\bibitem{Bruno01} \textit{Bruno A.D.} Power geometry in algebraic and differential equations,
Moscow, Nauka, Fizmatlit, (1998), 288 p (in Russian).

\bibitem{Bruno02} \textit{Bruno A.D.} Asimptotics and expansions of solutions of ordinary differential equations,
Uspehi of mathematical nauk, v. 59, No. 3, (2004), p. 31-80 (in
Russian).

\bibitem{Kuramoto01} \textit{Y. Kuramoto, T. Tsuzuki}, Prog. Theor. Phys., \textbf{55}, (1976),
356

\bibitem{Sivashinsky01}\textit{G.I. Sivashinsky}, Physica D, \textbf{4}, (1982),
227 -- 235

\end{thebibliography}
\end{document}